\documentclass[prb,twocolumn,superscriptaddress,showpacs]{revtex4}

\usepackage{times}
\usepackage{amsmath}
\usepackage{amssymb}
\usepackage{graphicx}
\usepackage[noxcolor]{pstricks}

\begin{document}

{}\vspace*{2cm}

\title{Collective excitations and universal broadening of cyclotron absorption in Dirac semimetals in a quantizing magnetic field}  

\author{D. I. Yasnov}  
\affiliation{Nizhny Novgorod State University, 603950 Nizhny Novgorod, Russia}
\affiliation{Institute of Applied Physics of the RAS, 603950 Nizhny Novgorod, Russia} 
\author{A. P. Protogenov}  
\affiliation{Nizhny Novgorod State University, 603950 Nizhny Novgorod, Russia}
\affiliation{Institute of Applied Physics of the RAS, 603950 Nizhny Novgorod, Russia} 
\author{P. M. Echenique}
\affiliation{Departamento de F\'isica de Materiales UPV/EHU, Centro de F\'sica de Materiales CFM - MPC and Centro Mixto CSIC-UPV/EHU, 20080, San Sebastian/Donostia, Spain}
\affiliation{Donostia International Physics Center (DIPC), 20018, San Sebastian/Donostia, Spain}
\author{E. V. Chulkov}
\affiliation{Departamento de F\'isica de Materiales UPV/EHU, Centro de F\'sica de Materiales CFM - MPC and Centro Mixto CSIC-UPV/EHU, 20080, San Sebastian/Donostia, Spain}
\affiliation{Donostia International Physics Center (DIPC), 20018, San Sebastian/Donostia, Spain}

\begin{abstract}
The spectrum of electromagnetic collective excitations in Dirac semimetals placed in a quantizing magnetic field is considered. 
We have found the Landau damping regions using the energy and momentum conservation law for allowed transitions between one-particle states of electron excitations. Analysis of dispersion equations for longitudinal and transverse waves near the window boundaries in the Landau damping regions reveals different types of collective excitations. We also indicate the features of universal broadening of cyclotron absorption for a magnetic field variation in systems with linear dispersion of the electron spectrum. The use of the obtained spectrum also allows us to predict a number of new oscillation and resonance effects in the field of magneto-optical phenomena.   

\end{abstract}

\pacs{71.55.Ak, 71.90.+q, 74.25.Jb, 71.70.Di, 52.27.Ny.}
\maketitle

\section{Introduction}

The existence of massless low-energy electron states \cite{First,Vol} in Dirac materials is one of the most important properties of these media. The massless excitations in Dirac semimetals are chiral as in graphene and are located inside the Brillouin zone in contrast to the graphene case. The stability of this  electron phase state with respect to weak structural disorder is guaranteed by the nonzero Chern topological invariant. The study of the absorption of electromagnetic waves in semimetals in a high magnetic field is one of the main methods for exploring resonance phenomena in such materials. A fundamental constraint in this field of consideration is imposed by the Kohn theorem \cite{Kohn} which states that the cyclotron frequency is the only frequency near which absorption occurs in infinite systems with a quadratic dependence of the energy of electrons on the momentum. It is important that this statement is valid with allowance for electron-electron interaction. If the dispersion law of electrons differs from quadratic, deviations from the Kohn law are minimal. In this paper, we find conditions under which deviations from the Kohn law can be significant. 
To achieve this task, we consider a Dirac semimetal placed in a quantizing magnetic field. Chiral electronic states with a linear dispersion law of low-energy electronic excitations exist in three-dimensional Dirac semimetals where inversion or time reversal symmetry is broken. One of these symmetries is broken in a number of crystals, e.g., CdAs, ZrTe, and ZrBeSi and $A_{3}B$ families, where A = (Na, K, Rb) and B = (As, Sb, Bi). The properties of Dirac semimetals are analyzed in detail in the review in Ref. \cite{Arm}. 

In order to study resonance properties of a Dirac semimetal in a quantizing magnetic field, we, for simplicity, consider the $Cd_{3}As_{2}$ compound whose Dirac spectrum of low-energy electronic states was explored in Refs. \cite{Bor,Liu,Neu}. To study the cyclotron absorption of electromagnetic waves, we need exact expressions for the spectrum and wavefunctions of Dirac electrons in the quantizing magnetic field. Their application in conservation laws of energy, momentum, and angular momentum of electronic excitations involved in the absorption of a photon together with the Pauli exclusion principle makes it possible to find the Landau collisionless damping region for 
longitudinal and transverse collective excitations. 

We consider longitudinal and transverse left-hand circularly polarized electromagnetic waves propagating along the magnetic field at an arbitrary frequency  and an arbitrary wave vector  of the electromagnetic field. Near the wave vector q=0 at finite frequencies and for filling of one to five Landau levels, we detected a significant broadening of cyclotron absorption. This phenomenon is completely caused by the existence of a "massive" spectrum of relativistic Dirac electrons in the quantizing magnetic field. In this case, a significant difference arises between the frequencies of electron transitions between states with Fermi momenta  corresponding to the neighboring n-th and (n + 1)-th Landau levels. In the last Section, we discuss some magneto-optical phenomena where the found effect occurred.

Massless electron excitations form a specific spectrum of collective modes in Dirac semimetals. The dispersion of collective modes in these Dirac materials in the absence of a magnetic field has been studied in many papers \cite{S,HS,RK,YHW,LvZ,Po,KL}\,. 
In particular, the plasma frequency in such media is inversely proportional to the square root of the Planck constant \cite{S}\,. The spectrum of surface plasmon polaritons in Weyl semimetals in a magnetic field was the subject of Ref. \cite{HoSa}\,. When the electrons at the Fermi energy share only the zero Landau level,  the unusual physical properties of Weyl semimetals in these extremely strong magnetic fields result in a collective mode \cite{SS,SA} with the finite frequency in the long-wavelength limit. This phenomenon is reflected in the response functions calculated for Dirac cones with allowance for the chirality sign \cite {SS,SA,ZCX}\,. 

In this paper, we consider Dirac semimetals placed in a quantizing magnetic field on the condition that the Landau level index at the Fermi energy is finite. The Landau level index lists the emerging quasi-one-dimensional subsystems that participate in relative motions with respect to each other. Such quasi-neutral oscillations of one-particle subsystems can form new branches in the spectrum of collective excitations. Collective modes will be undamped if their spectra belong to windows in collisionless damping regions. Therefore, we will start with finding the Landau damping regions of longitudinal and transverse collective excitations at arbitrary frequencies and wave vectors. 

The key idea in the study of the spectrum of collective modes is as follows. In contrast to systems with parabolic electron dispersion, contributions to the energy of longitudinal and transverse motion with respect to the direction of the magnetic field (taking into account the square root in the relativistic case) are not additive. It will be shown below that this gives rise to a novel type of collective excitations. 

\section{Longitudinal collective excitations}
To simplify the analysis, we assume that collective excitations propagate 
in the $z$ direction parallel to the magnetic field. The spectrum of longitudinal collective excitations is determined by the solutions of the dispersion equation  $\varepsilon_{l}(q_{z}, \omega) =0$. The dielectric function   
\begin{equation}\label{e1:example}
\varepsilon_{l}(q_{z}, \omega)=1-
V_{0}
\Pi(q_{z}, \omega)
\end{equation}
includes the Fourier transform $V_{0}$ of the bare Coulomb interaction 
and the polarization operator $\Pi(q_{z}, \omega)$ that depends on the wave vector $q_{z}$ directed along the magnetic field and the frequency $\omega$. 

The polarization operator in the considered Dirac spectrum and quantizing magnetic field in the  random phase aproximation has the form \cite{S}  
\begin{widetext}
\begin{equation}
\label{e2:example}
\Pi(q_{z}, \omega)=\frac{1}{2\pi l_{H}^{2}}\sum_{n, n^{\prime}}\int \frac{d p_{z}}{2\pi\hbar} |M_{n,n'}|^{2}\frac{f_{0}(E_{n}^{+}(p_{z})) - f_{0}(E_{n'}^{+}(p_{z}+\hbar q_{z}))}{\hbar\omega +E_{n}^{+}(p_{z})-E_{n'}^{+}(p_{z}+\hbar q_{z})+i0^{+}} \, .
\end{equation}
\end{widetext}

In Eq. (\ref{e2:example}),  
the summation over $n, n'$ is taken over all occupied Landau levels, $n=0,\pm 1,\pm 2,... $, $n_{F}$ is the maximum Landau level index, $E_{n}^{\pm}(p_{z})=\pm\sqrt{v_{F}^{2}p_{z}^{2}+\epsilon_{0}^{2}|n|}$  is the spectrum of massless Dirac electrons in the magnetic field $H$ \, \cite{Gavr}\,, $\epsilon_{0}=\sqrt{2}\hbar v_{F}/l_{H}$, $l_{H}=(c\hbar/eH)^{1/2}$ is the magnetic length, $p_{z}$ is the electron momentum along the magnetic field, $v_{F}$ is the Fermi velocity determining the slope of the Dirac cone, $c$ is the velocity of light, $e$ is the elementary charge, $\hbar$ is the Planck constant, and  $f_{0}(E_{n}^{\pm}(p_{z}))$ is the equilibrium Fermi distribution function at zero temperature. The case with finite temperature and finite lifetime of electrons due to collisions will be examined in the last section. The matrix element $M_{n,n'}$ is nonzero for the transitions with $\Delta n=n'-n=0$ for longitudinal waves. 

The formula for polarization operator (\ref{e2:example}) takes into account only the interband transitions for the restricted region of the frequency $\omega$ and the wave vector $q_{z}$. We will limit ourselves to considering such values of frequencies and the wave vectors. 
For circularly polarized waves, the angular momentum selection rules are such that the corresponding matrix element is nonzero for $\Delta n=n'-n=+1$ for left-hand circularly polarized waves and for $\Delta n=n'-n=-1$ for right-hand circularly polarized waves \cite{DP}\,. We do not give here the  matrix elements, because they are not essential for constructing the regions of Landau damping of longitudinal  waves (see Ref. \cite{DP})\,. 

Collisionless absorption of collective excitations due to their damping at electron-hole excitations are described by the pole bypass rule $i0^{+}$ in Eqs.  
(\ref{e2:example}). The regions of collisionless damping of collective excitations are related to the excitations of electron-hole pairs and are determined by the nonzero imaginary component of the dielectric function [Eq. (\ref{e1:example})]. Using the energy and momentum conservation law for this process 
\begin{equation}\label{e3:example}
E_{n}^{+}(p_{z})+\hbar\omega=E_{n'}^{+}(p_{z}+\hbar q_{z})
\end{equation}
and the conditions 
\begin{equation}\label{e4:example}
   E_{n}^{+}(p_{z}) \le E_{F}, \,\,\,\,\,\,
   E_{n'}^{+}(p_{z}+\hbar q_{z}) \ge E_{F} \, ,
\end{equation}
where $E_{F}$ is the Fermi energy, we obtain the Landau damping regions for excitations in Dirac semimetals placed in a quantizing magnetic field. For brevity, we introduce the dimensionless variables ${\cal{E}}_{n}^{+}(k_{z})=E_{n}^{+}(p_{z})/\epsilon_{0}$, 
${\cal{E}}_{F}=E_{F}/\epsilon_{0}$, $k_{z}=v_{F} p_{z}/\epsilon_{0}$, $\Omega= \hbar \omega /\epsilon_{0}$, and $q=l_{H}q_{z}/\sqrt{2}$.  
The solutions of inequalities (\ref{e4:example}) are a set of inequalities for longitudinal $(\Delta n =0)$, transverse left-hand circularly polarized $(\Delta n =1)$, and right-hand circularly polarized $(\Delta n =-1)$ waves. 

In this section, we consider longitudinal excitations. We provide below the exact boundaries of the Landau damping regions of longitudinal waves in Dirac semimetals placed in a quantizing magnetic field. For longitudinal waves $(\Delta n =0)$ shown in Figs. 1, the Landau damping regions for $0 \le \Omega\le 1$ and $0 \le q \le 1$ are determined by the inequalities
\begin{equation}\label{Apend1}
 \begin{cases}
   \Omega \le -{\cal E}_{F}+\sqrt{{\cal E}_{F}^{2} +q^{2} +2q \sqrt{{\cal E}_{F}^{2}-n}} \, , 
   \\
   \Omega \ge |-{\cal E}_{F}+\sqrt{{\cal E}_{F}^{2} +q^{2} -2q \sqrt{{\cal E}_{F}^{2}-n}}| \,. 
 \end{cases}
\end{equation}
The corresponding Landau damping regions are shown in Fig. 1. 
\begin{figure}[t]
\begin{center}
\includegraphics[width=0.8\columnwidth]{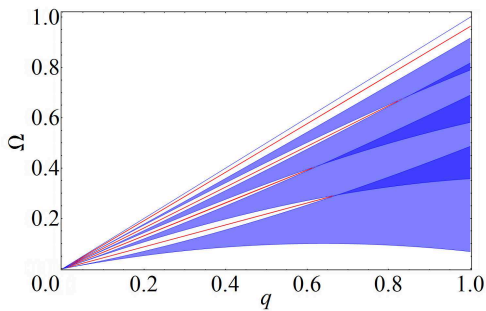} 
  \caption{(Color online). Landau damping regions for longitudinal waves for $\Omega \ll {\cal{E}}_{F}, q \ll k_{F}$, and $n_{F}=4$. The dispersion of acoustic magnetoplasmon excitations is plotted by the red lines.}
\end{center} 
 \label{fig:Dens}
\end{figure}

The line $\Omega=q$ in the plane $(q, \Omega)$ corresponds to the zero Landau index $n$. 
The windows in the Landau damping regions can be conventionally divided into two groups: the windows adjacent to the axis $\Omega$ and the windows adjacent to the axis $q$. Without a magnetic field, the frequency  
$\omega_{p}=
\sqrt{e^{2}/\hbar v_{F} }\left(64\pi /3\right )^{1/6}v_{F}n^{1/3}_{b}$
of longitudinal plasma oscillations \cite{S} in the long-wave limit and for $e^{2}/\hbar v_{F} =1$ 
is located in the first-type transparency region. Here $n_{b}$ is the electron bulk  density. 
This also takes place in the limit of a very strong magnetic field when $n_{F}=0$ and $\omega_{p}=
\sqrt{e^{2}/\hbar v_{F} }\left(2/\pi \right )^{1/2}v_{F}/l_{H}$ \cite{SS,SA}\,. 

The vanishing of Landau damping at the points 
$q_{n}=2\sqrt{{\cal{E}}_{F}^{2} -n}$ along the axis $q$ is reflected in singularities of the real part of the polarization operator near these points, which leads to oscillations of the static screening potential with an oscillation period determined by the position of these points. 
In the regions in the plane $(q, \Omega)$, there are groups of transparency windows in the Landau damping regions, the positions of which are given by discrete values $v_{n}=v_{F}\sqrt{1-n/{\cal{E}}_{F}^{2}}$ of the Dirac electron velocity at Landau levels. 

It is seen that in the transparency windows located in collisionless damping regions at low frequencies $\Omega$ and for a small wave vector $q$, there can exist 
\cite{MM,GKP,KP} undamped collective excitations with linear dispersion $\Omega = V_{n}q$. In Fig. 1, we schematically show the dispersion of these longitudinal collective waves. Their velocity $V_{n}$ is in the range $(v_{n+1}, v_{n})$ of Dirac electron velocities at neighboring Landau levels. The existence of a series of these acoustic magnetoplasmon excitations as solutions of the dispersion equation $\varepsilon(q_{z}, \omega)_{l}=0$, which in the considered region has the form 
\begin{equation}\label{e4:PSplasmons}
\sum_{n=0}^{n_{F}} \frac{v_{n}}{v_{n}^{2}-V_{n}^{2}} = 0\, ,
\end{equation}
follows from the fact that the real part of the polarization operator has singularities at the boundaries of the transparency windows due to the Kramers-Kronig relations. 
Different signs of the singularities lead to the solutions of the dispersion equation due to the intersection of the singular part of the equation with its smoothly varying part in the considered frequency range. These collective modes are similar to weakly damped acoustic Pines-Schrieffer plasmons in a degenerate two-component plasma with significantly different carrier masses. For the relativistic dispersion of electron excitations, the spectrum of undamped collective modes terminates at different frequencies, in contrast to the nonrelativistic quadratic electron dispersion in a quantizing magnetic field \cite{MM,GKP,KP,DP}\,. 

If the Landau level index $n$ is equal to its maximum $n_{F}$, the electron velocity $v_{n}$ at this level for a certain magnetic field can coincide with the sound speed $s$. If the condition $v_{n}=s$ is violated, there is no collision damping (the dispersion dependence for sound is in the transparency window). For a varied magnetic field, this leads to giant oscillations of sound absorption by Dirac electrons at Landau levels. The difference of this effect from a similar phenomenon for a parabolic electron dispersion in a quantizing magnetic field consists in nonequidistant position of maxima of the sound absorption coefficient in its dependence on the magnetic field. This difference is due to the fact that $E_{n}^{+}(p_{z})\sim \sqrt {Hn}$ for $n >> 1$.  If the condition $s < v_{n}$ with $n=n_{F}$ is violated, sound is absorbed at discrete frequencies $\Omega_{n}$ corresponding to discrete values of the wave vector $q_{n}=\Omega_{n}/s$ (see Fig. 1). 

\section{Left-hand circularly polarized collective excitations}

Let a circularly polarized electromagnetic wave with the frequency $\omega$ and wave vector $q_z = q$ propagate along a magnetic field with the strength $H$. 
The dispersion equation $\varepsilon (\omega,q)_{+}=c^{2}q^{2}/\omega^{2}$ for left-hand  circularly polarized transverse electromagnetic modes includes the dielectric function. 
In the random phase approximation with the small parameter $r_s = \frac{e^{2}}{\varepsilon_{0}\hbar v_{F}}$, which is similar to the fine structure constant $\frac{e^{2}}{\hbar c}$, this function has the form 
$$
\varepsilon (\omega,q)_{+}=\varepsilon_{0}+\frac{\omega_{p}^{2}}{\omega \omega_{c}}\,, 
$$
\begin{widetext}
\begin{equation}
\label{eq:e1}
\frac{1}{\omega_{c}}=\hbar^{2}v_{F}\sum_{n n'ss'}^{n_{F}} \int dk_{z} \frac{f_{0}(E_{n,s}(k_{z})) - f_{0}(E_{n',s'}(k_{z}+q))}{E_{n,s}(k_{z})-E_{n',s'}(k_{z}+q)}
\frac{|cos \theta_{n',s'}(k_{z}+q)\cdot sin \theta_{n,s}(k_{z})|^{2}}{E_{n,s}(k_{z}) + \hbar \omega - E_{n',s'}(k_{z}+q) +i0^{+}} F_{ss'}(k_{z},k_{z}+q). 
\end{equation}
\end{widetext}

Here, $\varepsilon_{0}$ is the background lattice dielectric constant of the system, and  
$\omega_{p}^{2}=\frac{e^{2}}{\hbar v_{F}} \frac{2}{\pi} (\frac{v_{F}}{l_{H}})^{2}$ is the square of the plasma frequency in the case of filling of the zeroth Landau level \cite{SS,SA} at $\varepsilon_{0} = 1$;  summation over the Landau level numbers $n$ and $n'$ is performed up to the extreme Fermi number $n_{F}$ of the Landau level,   
\begin{equation}
\label{eq:e3}
\tan \theta_{n,s}(k_{z})=\frac{E_{n,s}(k_{z}) -\hbar v_{F} k_{z}}{\epsilon_{0} \sqrt{n}} \, .
\end{equation} 
 
The subscripts $s = +1$ and $-1$ correspond to the conduction $ (+1)$ and valence $ (-1)$ bands, respectively. The presence of the function $F_{ss'}(k_{z},k_{z}+q) = (1 + ss'\cos \vartheta )/2$ ($\vartheta $ is the angle between $k_{z}$ and $k'_{z}=k_{z}+q$) in Eq. (\ref{eq:e1}) is similar to the situation in graphene and reflects the chiral nature of electrons in a Dirac semimetal, including the contribution of interband transitions, e.g.,, the  transition $(s=-1 \to s' = +1)$ \cite{LvZ} in addition to the standard $(s=+1 \to s' = +1)$ intraband transition. 
The chosen electron variant of filling of bands corresponds to the case where the Fermi energy 
is $E_{F} > 0$. 

We now discuss the collisionless Landau damping regions of left-hand circularly polarized 
collective excitations. The rule of bypass of the pole in (\ref{eq:e1}) at a nonzero numerator makes it possible to determine the collisionless damping region of collective excitations. 
According to the law of conservation of the energy $E_{n,s}(k_{z})  + \hbar \omega = E_{n',s'}(k'_{z})$, momentum $k'_{z}=k_{z}+q$, and angular momentum $n' = n \pm 1$ \cite{DPr}, inequalities $E_{n,s}(k_{z}) < E_{F}, E_{n',s'}(k'_{z}) > E_{F}$  following from the Pauli exclusion principle, describe the Landau damping region. In the case of interest $n' = n + 1$ occurring for left-hand circularly polarized electromagnetic waves \cite{DP}, the collisionless damping regions are shown in Fig. 2 at the dimensionless frequency $\Omega= \hbar \omega /\epsilon_{0}$ and for dimensionless wave vector $Q=l_{H}q_{z}/\sqrt{2}$.

\begin{figure}[t]
\begin{center}
\includegraphics[width=0.8\columnwidth]{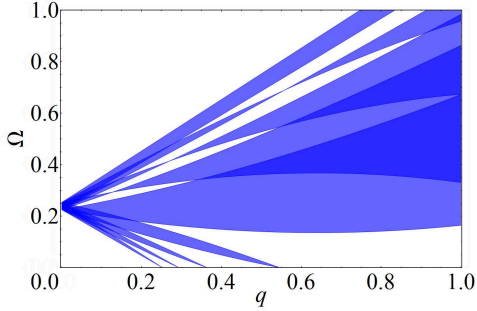} 
  \caption{(Color online). Landau damping regions for left-hand circularly polarized waves for $\Omega \ll {\cal{E}}_{F}, q \ll k_{F}$, and $n_{F}=4$.}
\end{center} 
 \label{fig:state}
\end{figure} 

For left-hand circularly polarized waves $(\Delta n =+1)$, the Landau damping boundaries for $0 \le \Omega\le 1$ and $0 \le q \le 1$ in Fig. 2 are determined by the following inequalities  
\begin{equation}\label{Ape5}
    \Omega \le -{\cal E}_{F}+\sqrt{{\cal E}_{F}^{2} +q^{2} +1 +2q \sqrt{{\cal E}_{F}^{2}-n }},   
    \, \,\, \,  
  q_{n}^{*}<q ,
   \\
 \end{equation}  
  \begin{equation}\label{Apend52} 
    \Omega \ge {\cal E}_{F}-\sqrt{{\cal E}_{F}^{2} +q^{2} -1 -2q \sqrt{{\cal E}_{F}^{2}-n -1}}, 
  q_{n}^{*}<q,
\end{equation}
\begin{equation}\label{Apend6}
\Omega \ge -{\cal E}_{F}+\sqrt{{\cal E}_{F}^{2} +q^{2} +1 +2q \sqrt{{\cal E}_{F}^{2}-n }} \, , \,\, \,\,\, 
    q_{n}^{*}>q ,
\end{equation}
  \begin{equation}\label{Apend62} 
  \Omega \le {\cal E}_{F}-\sqrt{{\cal E}_{F}^{2} +q^{2} -1 -2q \sqrt{{\cal E}_{F}^{2}-n-1}} \, , \,\, \, q_{n}^{*}>q ,
\end{equation}
\begin{equation}\label{Apen7}
 \Omega \le {\cal E}_{F}-\sqrt{{\cal E}_{F}^{2} +q^{2} -1 +2q \sqrt{{\cal E}_{F}^{2}-n-1 }} , \,\,\,\,\,\, \, q<q'''_{n} ,
  \end{equation}
 \begin{equation}\label{Apen72} 
  \Omega \ge -{\cal E}_{F}+\sqrt{{\cal E}_{F}^{2} +q^{2} +1 -2q \sqrt{{\cal E}_{F}^{2}-n}} ,  \,\,\, \, \,\, \, \,\,  q<q'''_{n} ,
\end{equation}
where \\
$q'''_{n}=\sqrt{ 2{\cal E}_{F}^{2} -2n -1 -2\sqrt{{\cal E}_{F}^{4}-{\cal E}_{F}^{2} -2 {\cal E}_{F}^{2} n +n^{2} +n}}$, and $q_{n}^{*}$ is  the point of the contact of transparency windows corresponding to different $n$. 

As distinct from the Landau damping regions for longitudinal waves, the regions for left-hand circularly polarized waves have a more complicated structure. 
This property substantially manifests itself even in the long-wavelength limit shown in Fig. 2.
In particular, for $q=0$, on the axis of the frequency $\Omega$, there is a frequency range $\Omega_{min}<\Omega < \Omega_{max}$ where collisionless damping is nonzero. Here $\Omega_{min}=-{\cal{E}}_{F} + \sqrt{{\cal{E}}_{F}^{2}+1}$ and  $\Omega_{max} ={\cal{E}}_{F} - \sqrt{{\cal{E}}_{F}^{2}-1}$. The lower boundary of the range, i.e., the frequency $\Omega_{min}$ arises as the energy difference between neighboring Landau levels. For $\Omega < \Omega_{min}$ and $q <  q'''_{n_{F}}$,  we find a region without collisionless damping of collective excitations.  
To clarify the physical meaning of the frequency range [$\Omega_{min},\Omega_{max}$], consider the electron energies ${\cal E}_{n}(k_{z})$ and ${\cal E}_{n+1}(k_{z})$ at two neighboring Landau levels for the momentum $k_{z}$ equal to $k_{F}^{n+1}$ and $k_{F}^{n}$.
Then the vertical transition frequency ${\cal E}_{n}(k_{F}^{n+1}) \to {\cal E}_{n+1}(k_{F}^{n+1})$ is $\Omega_{max}$, while the transition frequency ${\cal E}_{n}(k_{F}^{n})\to {\cal E}_{n+1}(k_{F}^{n})$ is $\Omega_{min}$. These frequencies are independent of the Landau level index. For a quadratic electron dispersion, these frequencies coincide and are equal to the cyclotron frequency. 

In the low-frequency range in the plane $(q, \Omega)$, there is a collective excitation in a strong magnetic field known as the helicon whose spectrum in Weyl semimetals was recently studied in Ref. \cite{IK}. This low-frequency mode with quadratic dispersion for increasing frequency and wave vector for its finite discrete values ${\bar q}_{n}$ is subjected to a collisionless Landau damping. If the magnetic field varies when passing from one transparency window to another, giant oscillations of the helicon damping coefficient occur. 

A new phenomenon arises in the region $\Omega > \Omega_{max}$. In this part of the plane $(q, \Omega)$ inside the collisionless damping regions, there are transparency windows in which novel collective excitations with finite values of the wave vector $Q$ can propagate. They are somewhat similar to the acoustic magnetoplasmon excitations considered above for waves with longitudinal polarization (see Fig. 3).  
\begin{figure}[t]
\begin{center}
\includegraphics[width=0.9\columnwidth]{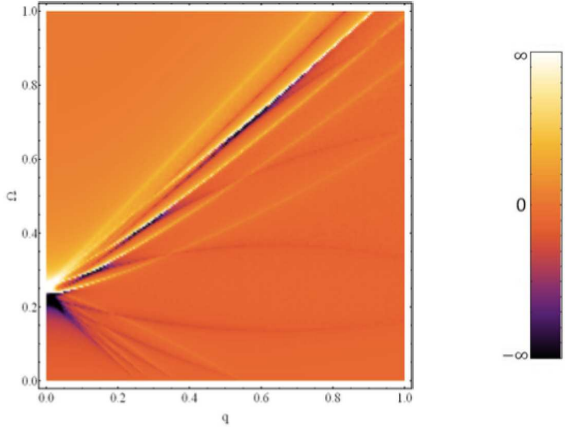} 
  \caption{(Color online). Numerical solution to the dispersion equation for the left-hand circularly polarized collective excitations. $n_{F}=4$.}
\end{center} 
 \label{fig:Dens1}
\end{figure}

We note that the windows in the collisionless Landau damping regions and the collective excitations are due to the absence of energy additivity mentioned in the introduction when the energies of longitudinal and transverse motions in a quantizing magnetic field are added under the square root of the relativistic dispersion.  Physically,  
the existence of the excitations is based on the possibility of longitudinal oscillations of two subsystems relative to each other, which are accompanied by rotation during the cyclotron period.

\section{Universal Broadening of Cyclotron Absorption} 

Collisionless damping regions at $Q=0$ in Fig. 1 include a remarkable frequency range $\Omega_{min} \leq  \Omega   \leq \Omega_{max}$ , where the absorption of electromagnetic waves occurs. The physical reason for the appearance of this range with the width $\Delta \Omega = \Omega_{max} - \Omega_{min}$ was discussed in the Introduction. The boundaries of the range depend on the dimensionless Fermi energy ${\cal E}_{F}=E_{F}/\epsilon_{0}$ as $\Omega_{max} = {\cal E}_{F} - \sqrt{{\cal E}_{F}^{2} - 1}$  
and $ \Omega_{min} = -{\cal E}_{F} + \sqrt{{\cal E}_{F}^{2} + 1}$. 

To describe the conditions of absorption of electromagnetic waves with frequencies above the plasma frequency, we use the relations $n_{F} = {\cal E}_{F}^{2} = l_{H}^{2}k_{F}^{2}/2 = H_{0}/H \equiv x^{2}/2$, where $k_{F} = (3\pi^{2}n_{b})^{1/3}$, $n_{b}$ is the bulk electron density, and $H_{0} = k_{F}^{2}c\hbar/(2e)$. 
To solve the problem of the transmission (absorption) of an electromagnetic wave through the plasma medium under consideration, it is necessary to know the magnetic field dependence of the reference plasma frequency $\overline{\omega}_{p}(x) = {\omega}_{p}/({\sqrt \varepsilon_{0}}v_{F}k_{F})$, where $x = \sqrt{2H_{0}/H }$. 
For an arbitrary magnetic field, this dependence is expressed in terms of the well-known asymptotic value of the plasma frequency $\omega_{pBS} = \sqrt{2r_{s}/\pi}v_{F}/l_{H}$ for a high magnetic field \cite{SS,SA} when only the zeroth Landau level is filled and the asymptotic value $\omega_{pDS} = \sqrt{r_{s}}(64\pi)^{1/6}v_{F}n_{b}^{1/3}$ in zero magnetic field \cite{S,HS}.  

The dimensionless cyclotron absorption edges at $Q=0$, which are introduced by the formula 
$\overline{\omega}=\omega/(v_{F}k_{F})$, and the dimensionless plasma frequency 
$\overline{\omega}_{p}=\omega_{p}/(v_{F}k_{F})$  as functions of the parameter $x$ are given by the expressions
\begin{equation}
\label{eq:e4}
\overline{\omega}_{max} = 1 - \sqrt{1 - 2/x^{2}} \, , 
\end{equation}
\begin{equation}
\label{eq:e5}
\overline{\omega}_{min} = -1 + \sqrt{1 + 2/x^{2}} \, , 
\end{equation}
\begin{equation}
\label{eq:e6}
\overline{\omega}_{p} = \frac{a}{x}\sqrt{1 + f(x)} \, . 
\end{equation} 
Here, $a=\sqrt{2r_{s}/\pi}$ and the function 
$$f(x)=\sum_{n=1}^{n_{F}} \sqrt{1-n/(x^{2}/2)} = x^{2}/3$$ at $n_{F} \gg 1$ and $f(\sqrt{2})=0$ at $n_{F}=1$. 

The dispersion dependence of transverse waves at 
$ \overline{\omega} > \overline{\omega}_{p}$ and $Q \to 0$ in the dimensionless variables has the form 
\begin{equation}
\overline{\omega}^{2}(Q)= \overline{\omega}_{p}^{2} + 
\frac{2c^{2}}
{\varepsilon_{0}v_{F}^{2}x^{2}} Q^{2} \, .
\end{equation}

\begin{figure}[t]
\begin{center}
\includegraphics[width=0.9\columnwidth]{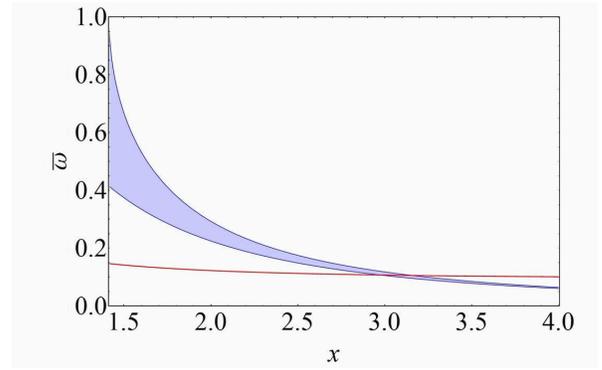}
\caption{(Color online). Frequencies (upper line) $\overline{\omega}_{max}$, (middle line) 
 $\overline{\omega}_{min}$, and (lower line) $\overline{\omega}_{p}$ at $a=0.16$ versus 
 the parameter $x$. Collisionless Landau cyclotron damping exists in the shaded region $\overline{\omega}_{min} < \overline{\omega} < \overline{\omega}_{min}$ between the upper and middle lines.}
\end{center} 
 \label{fig3:Lines}
\end{figure} 

The parameters given by Eqs. (\ref{eq:e4}) - (\ref{eq:e6}) are shown in Fig. 4 as functions of the parameter $x$ at $x \geq \sqrt{2}$. 
At $\sqrt{2} \leq  x < x_{min}$, the external electromagnetic wave with a fixed frequency $\overline{\omega} < \overline{\omega}_{min}$ will propagate without absorption at a decrease in the magnetic field to the value $x = x_{min}$ to the collisionless damping region from the frequency range $\overline{\omega}_{p}  < \overline{\omega} < \overline{\omega}_{min}$. 
With a further decrease in the magnetic field, the electromagnetic wave will again propagate without damping at $x > x_{max}$, when the frequency of external radiation $\overline{\omega} = const$, being above the plasma frequency $\overline{\omega}_{p}$, intersects the curve 
$\overline{\omega}_{max}$ in Fig. 4. 

The width $\Delta \overline{\omega}$ of the collisionless cyclotron damping band depends on the magnetic field $H$ and electron density $n_{b}$. As is seen in Fig. 4, this width is maximal at $n_{F}=1$, when $\Delta \overline{\omega}=2-\sqrt{2}$, and decreases at an increase in the number of filled Landau levels ($\overline{\omega}_{max} \approx \overline{\omega}_{min} \lesssim  \overline{\omega}_{p}$  at $n_{F}=5$). We have considered the case where the external magnetic field is a variable parameter. A similar change of the regime of transmission of the external electromagnetic field to the regime of its absorption with the subsequent possibility of transmission occurs if the frequency of the electromagnetic field is a control parameter at 
$x=const$. 

\section{Discussion}  

In this paper, we have analyzed the effect of collisionless Landau damping in Dirac semimetals in a quantizing magnetic field at the propagation of electromagnetic waves in them. Since the electron spectrum in such systems is nonequidistant, new transparency windows appear in the Landau damping regions at $q \sim k_{F}$, where new collective modes can exist, and the cyclotron absorption at $q \approx 0$ is broadened. 
Electromagnetic waves in systems placed in the quantizing magnetic field at $q \approx 0$  are usually absorbed at the cyclotron frequency. In systems whose electron spectrum is nearly square, this occurs under wider conditions \cite{Kohn}. In Dirac semimetals with a relativistic dispersion law in the ultraquantum case where a few Landau levels are filled, the absorption of electromagnetic waves occurs in the cyclotron frequency range $[\omega_{min},\omega_{max}]$. 
The width of the range depends on the number of filled Landau levels. 

In the discussion of the Landau damping and spectrum of collective modes, we assume that temperature dependence of the distribution function is absent and the lifetime of electron excitations is the largest of all time scales. When taking into account the final temperature $T$ and the final lifetime $\tau$ of the charge carriers, these collective phenomena can occur if $k_{B}T \ll  \epsilon_{0}(\epsilon_{0}/E_{F})$ and $\hbar/\tau \ll  \epsilon_{0}(\epsilon_{0}/E_{F})$ because the right-hand side of these conditions is the difference $E_{n+1}(k_{Fn})-E_{n}(k_{Fn}) = \sqrt{E_{F}^{2}+\epsilon_{0}^{2}} - E_{F}
\approx \epsilon_{0}(\epsilon_{0}/2E_{F})$ between the Landau levels for $k_{Fn}=\sqrt{E_{F}^{2} - \epsilon_{0}^{2}|n|}/\hbar v_F$. Here $k_{B}$ is the Boltzmann constant and $E_{F}$ is the Fermi energy. These conditions can be provided at helium temperatures and in magnetic fields of the order of $40$ T. 

Standard conditions for the experimental observation of cyclotron resonance $\nu \ll \omega_{c}$ and $k_{B}T \ll  \hbar\omega_{c}$ hold in our case. If the frequency dependence of damping in the frequency range $\Delta\omega=\omega_{max}-\omega_{min}$ is of interest, the width of the frequency range $\Delta\omega=\omega_{max}-\omega_{min}$ rather than cyclotron frequencies should be used in these inequalities for an estimate. This is necessary to avoid “smearing” of the frequency range $[\omega_{min},\omega_{max}]$  because of broadening of energy levels associated with a finite lifetime of electron states and temperature expansion of the distribution function. The modified observation conditions should now have the form $\nu \ll \Delta\omega$  and $k_{B}T \ll  \hbar\Delta\omega$. The dimensional frequencies 
$\omega_{min}$ and $\omega_{max}$ can be represented as  
\begin{equation}
\label{eq:omega1}
\omega_{min}=-\frac{E_{F}}{\hbar} + \frac{v_{F}}{l_{H}} \sqrt{\frac{E_{F}^2 l_{H}^2}{\hbar^2 v_{F}^2}+2}, 
\end{equation}

\begin{equation}
\label{eq:omega2}
\omega_{max}=\frac{E_{F}}{\hbar} - \frac{v_{F}}{l_{H}} \sqrt{\frac{E_{F}^2 l_{H}^2}{\hbar^2 v_{F}^2}-2}.
\end{equation}

For the parameters $v_{F}=1.5 \cdot 10^8$ cm/s and $E_{F}=0.25$ eV ($n_{b}=6 \cdot 10^{17}$ cm$^{-3}$) and external magnetic field $B=5$ T ($n_{F}=4$), the condition $k_{B}T \ll  \hbar\Delta\omega$ is satisfied at $T=10$ K. In this case, $k_{B}T=0.0008$ eV 
and $\hbar\Delta\omega = 0.004$ eV. The boundaries and width of the frequency range are 
$\omega_{min}=4.4 \cdot 10^{13}$ s$^{-1}$, $\omega_{max}=5.1 \cdot 10^{13}$ s$^{-1}$, 
and $\Delta\omega=0.7 \cdot 10^{13}$ $s^{-1}$, respectively. 
 
The frequencies $\omega_{min}$  and $\omega_{max}$ become dependent on $q$ in 
the region $q \sim k_{F}$. In terms of the dimensionless variables $Q$ and $\Omega$ 
introduced above, these dependences have the form 
\begin{equation}
\label{eq:OmQ1}
\Omega_{min}=-{\cal{E}}_{F}+\sqrt{2 Q \sqrt{{\cal{E}}_{F}^2 -n}+{\cal{E}}_{F}^2+Q^2+1}, 
\end{equation}

\begin{equation}
\label{eq:OmQ2}
\Omega_{max}={\cal{E}}_{F}- \sqrt{ -2 Q \sqrt{{\cal{E}}_{F}^2 -n-1} +{\cal{E}}_{F}^2+Q^2-1}.
\end{equation}

These expressions determine the edges of transparency windows; the conductivity singularities near these edges have different signs. The contribution of damping regions for interband transitions at $\omega \sim 2\omega_{min/max}$ is insignificant in the regions considered above ($\omega \approx \omega_{min/max}$). 

The above analysis reveals the following effects. The first effect is a change in the polarization of a circularly polarized electromagnetic wave with a variation of the magnetic field when its frequency appears in the collisionless cyclotron damping region where the propagation of the 
left-hand circularly polarized wave is suppressed. The second effect is the control of the considered effect by an electric field perpendicular to the magnetic field \cite{Gav1,Gav2}. 
In this case one has to use the pulse mode to avoid the electric field screening.
For experimental studies of the collective phenomena investigated in the present paper, one should employ a combination of high magnetic fields, low temperatures and appropriate frequencies as in the experiments \cite{Gavr}.

The third effect is related to the helicon spectrum which in the considered ultraquantum limit has the following form   
\begin{equation}
\label{eq:helicon}
\overline{\omega}=\frac{2\overline{\omega}_{c}c^{2}}{\overline{\omega}_{p}^{2}v_{F}^{2}x^{2}}Q^{2} \, .
\end{equation}
Here, $\overline{\omega}_{c}=b/x$ where the coefficient $b$ can be find by integration 
in Eq. (\ref{eq:e1}).

In summary, we have found longitudinal and transverse 
collective electromagnetic excitations in Dirac semimetals in a quantizing magnetic field. 
We have also described universal broadening of cyclotron absorption in systems with linear dispersion of the electron spectrum.

\section{Acknowledgments}

We are grateful to V. L. Bratman, V. V. Zheleznyakov, R. V. Turkevich, and S.V. Eremeev for stimulating discussions. This work was supported in part by Project No. 18-12-00169 of the Russian Science Foundation.

\end{document}